# Bypassing Cloud Providers' Data Validation to Store Arbitrary Data

Guilherme Sperb Machado, Fabio V. Hecht, Martin Waldburger, Burkhard Stiller
Department of Informatics IFI, University of Zurich
Binzmühlestrasse 14, CH—8050 Zurich, Switzerland
[machado | hecht | waldburger | stiller]@ifi.uzh.ch

*Abstract*—A fundamental Software-as-a-Service (SaaS) characteristic in Cloud Computing is to be application-specific; depending on the application, Cloud Providers (CPs) restrict data formats and attributes allowed into their servers via a data validation process. An ill-defined data validation process may directly impact both security (*e.g.* application failure, legal issues) and accounting and charging (*e.g.* trusting metadata in file headers). Therefore, this paper investigates, evaluates (by means of tests), and discusses data validation processes of popular CPs. A proof of concept system was thus built, implementing encoders carefully crafted to circumvent data validation processes, ultimately demonstrating how large amounts of unaccounted, arbitrary data can be stored into CPs.

## I. INTRODUCTION

Cloud Computing (CC) is a concept that has been hyped to extremes over the last couple of years. Different definitions, different characteristics, and different categories (*i.e.*, types of Clouds or types of Cloud Services) made CC a prominent technology to be explored from private to corporate users. As result, a wide variety of Cloud Services (CS) were released by several Cloud Providers (CP), *e.g.*, Amazon EC2 [2], Google Picasa [8], SoundCloud [15], amongst others. Cloud Services can be classified into three different types [3]: Infrastructure-as-a-Service (IaaS), Platform-as-a-Service (PaaS), and Software-as-a-Service (SaaS). SaaS is the highest level of CC, meaning that users do not need to be concerned on PaaS or IaaS in order to use software that runs in the Cloud. SaaS features a complete application offered as a service on demand.

One of the main fundamental SaaS characteristics is to be *application-specific*. As the name says, it is *software* made available as a *service*, with a specific purpose and use. Therefore, depending on the application itself, CPs restrict data pushed into their servers. Google Picasa is an example to illustrate SaaS restrictions related to what is pushed to the application. Within Picasa, nowadays, the user can upload photos but under certain limitations: just a set of image formats (*e.g.*, JPG, PNG, BMP), resolution up to 2048x2048 pixels (under free service plan), file size up to 20 MByte. In order to enable application-specific restrictions, CPs must implement a *data validation* process [21] to verify what is being uploaded to the SaaS application.

Hence, this paper investigates, evaluates (by means of tests), and discusses the data validation process of CPs in the scope of SaaS. The research value and innovation of the paper relies on answering the following research question: is it possible to bypass the data validation process to store arbitrary data into CPs' servers? If it is possible (based on the test results), observe the impacts related to (1) security as well as (2) accounting and charging. These two dimensions are key to CPs since reputation and revenue can be affected.

In the scope of (1), a poor or incorrect data validation process can lead to security vulnerabilities, depending on how the SaaS application is implemented and internally architected — even though this paper does not aim to expose application-specific vulnerabilities. Also, an input not recognized by the data validation process can allow data injection to CPs' servers bypassing content rules and persisting content that the provider is unaware of. Related to (2), the data validation process must be aware of consistency. First, it must assure that what is declared is actually used (or vice versa). *E.g.*, if an uploaded photo has resolution declared as 80x60 in the header, but the payload data size as 1000x1000 pixels, the image should not be valid. Second, it should be aligned to what the Accounting System (AS) takes into consideration. *E.g.*, if a CP does not account and charge based on storage, the data validation process should verify whether pushed data do not contain an abusive data amount considering allowed file formats. Thus, depending on how CPs implement the accounting of resources, a poor data validation process may impact to what users consume (accounting) and what they will have to pay for (charging). Based on (1) and (2), the discussion part of the paper focuses on: how malicious users can benefit of encoding *data inside data* (bypassing the data validation an content rules), how CPs can have economic losses (accounting/charging), and how CPs can be legally affected.

In order to investigate and evaluate the data validation process, a set of *encoders* and *test cases* were implemented and performed on well-known public CSs (*e.g.*, Google Picasa, Facebook, SoundCloud, TwitPic). The tests simulate a user uploading and retrieving data (*i.e.*, files) via a proof of concept software — developed in the context of this paper — that constructs files in a particular manner based on different *encoders*. Each encoder implements a different technique,

*e.g.*, using steganography [11], or exploring the use of specific header fields.

The remainder of this paper is organized as follows. Section II presents the terminology and related work. Section III describes the methodology for the tests. Section IV presents the encoders used in the test cases, and the proof of concept system to automate the tests. Section V describes the test cases design. Section VI presents the test results, while Section VII discusses observed impacts. Finally, Section VIII concludes what was achieved including future work.

## II. TERMINOLOGY AND RELATED WORK

For SaaS applications studied in the scope of this paper, the common denominator for users to interact with the Cloud (besides text input) are *computer files*. A *computer file* (or just *file*) is composed of *organizational data* and *content data*. The former is represented by headers or control information, usually following a certain standard. In more general terms, *organizational data* can also be called *metadata*, since it is "data about *content data*". The raw term *metadata* brings ambiguity due to different fundamental types, and, therefore, the term "*organizational data*" was chosen. The latter is the data which the *organizational data* describes.

A *file format* is a particular way that information is encoded for storage in a *computer file*, as files need a way to be represented as bits when stored on a digital storage medium [18]. *Encoding* is the process by which information from a source is converted into bits (considering a *file format*). *Decoding* is the reverse process, converting these bits back into information understandable by a receiver. Usually, a *computer file* has a *file name extension* that is a popular, intuitive, and human-readable method to identify a *file format*. It is represented by a *computer file* suffix separated from the file name by a dot symbol (*e.g.*, filename.mp3, filename.jpg). *File name extensions* can also be considered a type of metadata.

Related work has already explored the data validation process, but not in the context of Cloud Computing. *E.g.*, solely focusing on security, Brumley *et al.* [5] propose a technique to automatically generate exploits from software patches that target input data validation vulnerabilities. Yoon *et al.* [21] present a framework for the validation of data and rules in Knowledge Base systems. For the given systems, the framework defines the validation of data, validation of rules [9], and the interaction between data and rule validation. Gancheva *et al.* [7] proposed a data security and data validation framework for a SOA (Service Oriented Architecture) based system. The paper explains what are employed levels of data security: user authorization access, data encryption, and data validation. As stated in [7], Petukhov *et al.* [13] also stresses that the main causes for errors, vulnerabilities, and consequently serious impacts, are due to poorly or non-validated data. On the topic of injecting data in Cloud Providers, GMailFS [19] is a tool that builds a file system using Google GMail service. GMailFS allows the storage of files, independent of their type and size. This is accomplished by segmented email attachments. Even misusing the SaaS application purpose, this tool does not exploit data validation weaknesses. In contrast to this paper, GMailFS does not use techniques to mask injected content in order to bypass CPs' data validation.

Data validation should be investigated in the area of CC since public CPs are directly exposed to impacts due to the considerable number of users, which can freely attempt to push data to CSs. As far as the authors are aware of, no research work moved towards the investigation of data validation process in the area of CC, focusing on real CPs and CSs, performing tests, and mainly understanding its impacts. No research used techniques as steganography or additional optional headers in order to inject data into CPs' servers bypassing data validation and content rules. Moreover, this paper introduces the problem of ill-defined data validation in the scope of CC.

## III. METHODOLOGY

The analysis of the methodology employed to analyze CPs' data validation is based on tests carried out on selected CPs, with different CSs — the selection criteria is described later in this section. The tests are classified as *exploratory testing*, which is defined by Bach *et al.* [4] as "simultaneous learning, test design, and test execution". This means that the tester actively controls the design of test cases and its execution, using the gained information to design new and better tests. As far as the authors are aware of, no research ever directed test cases to CPs' data validation process. Therefore, the selection of CSs, the designed test cases, and the implemented encoders do not follow any previous results.

### A. Test Goals

The tests presented in this paper evaluate if the data validation process can be bypassed using specific techniques in order to inject arbitrary data into CPs' servers. The tests' goals are directed to two impact dimensions:

**Security.** Evaluate if a file can carry data (data inside data) which the CP is not aware of, since the data validation process and content filtering system cannot detect it. *E.g.*, encode a PDF inside an image file, which is a file format that is accepted by the CS. Thus, the PDF content cannot be seen by the CP itself, which may contain, non-authorized information.

**Accounting and Charging.** Evaluate if a file encoded in different ways, and successfully uploaded to a CS, can impact on how the CP accounts consumed resources and charges users. In this case, the goal is to check if a file can negatively or positively influence the amount of resources the user consumed — also observing if the CP is aware of such consumption. *E.g.*, encode a PDF inside an MP3 file, making the MP3 file very big but just with 10 seconds of audio. In this example, the user would benefit of storing an amount of data that is not compatible to 10 seconds of audio (even though the

MP3 is encoded using the best quality) and therefore taking advantage of storing data in CPs' servers. Considering that the user can store 120 minutes of audio in a free account, then it could be possible to deceive the accounting/charging system in order to host many audio files of 10 seconds — but storing a huge quantity of data.

### B. Test Method

The employed test method simulates users pushing and retrieving data of a SaaS application. After the data is pushed to the application, this method assumes that data flows directly through the data validation process. If the data gets validated by validation rules, it is accepted, persisted and the SaaS makes it available for retrieval. The reason to also retrieve the content is due to the necessity of checking if the data constructed (through the means of encoders, where it will be discussed in Section IV) was maintained or modified.

The method is illustrated in the following steps:

**Step 1.** For a particular Cloud Service, check what are the restrictions to push data to the application. If no restrictions are present, the data validation process will not be invoked. Therefore, the tests should consider encoders which may explore accounting and/or charging impacts. *E.g.*, Dropbox allows its users to push anything to their account, limiting only the total space used.

**Step 2.** Based on the restrictions, construct the data to be pushed in a particular manner using encoders (cf. Section IV). The data should be encoded in order to answer the research question made in Section I and based on the data restrictions observed in Step 1. This process is key for the tests, since encoders can explore detailed factors in the data validation process.

**Step 3.** Push the data to the SaaS application. The data should be pushed following the CS's API (Application Programming Interface).

**Step 4.** Check if the data was successfully pushed (accepted) and became available to be retrieved. If the data was not accepted, the data validation process possibly detected a rule violation. Therefore, the CS is *not* susceptible to the employed encoder constructing the data (*unsuccessful* test result). Note that in blackbox testing it is impossible to observe the exact reasons for the denial. If the data was accepted, the data validation process is susceptible to the employed data encoder.

**Step 5.** In case of a successful push, retrieve the data that was pushed in the previous step.

**Step 6.** Based on the retrieved data: if the data is modified from its original encoding (prior to the step 3) the data validation process has a mechanism to reorganize the data preventing that malicious users inject data to retrieve it in the original format. Therefore, the data validation process is *not* susceptible to the employed encoder — being impossible to re-construct the original data again. If the retrieved data remained the same compared to the step before encoding, the test case is considered *successful*. Moreover, if the CP wrongfully accounted consumed resources — according to what the CP accounts and charges for — its data validator led the accounting system to wrong values.

All these steps are key to evaluate the data validation process. However, Steps 1, 3 and 5 are very service-specific and therefore require a particular effort for each tested CS.

### C. Cloud Provider's Selection Criteria

The CPs and CSs are selected based on popularity. It follows the logic that widely popular CPs tend to present a better-defined data validation process, while, in comparison, less popular CPs tend to present a less well-defined process. This paper assumes a higher probability that data validation process of popular CPs are refined over time — due to the high amount of data being pushed, and, therefore, more malformed data reported as not valid. In this scope, popularity of CSs is empirically measured observing, *e.g.*, number of users, volume of data pushed, and Alexa rank [1].

## IV. ENCODERS AND PROOF OF CONCEPT SYSTEM

Encoders are software implementations that construct files (data) in a particular manner to test CPs' data validation. Encoders are divided into three groups: Steganography-related (Section IV-A), FileFormatHeader-related (Section IV-B), and Appending-related (Section IV-C). The encoder groups are not only meant to sort similar techniques, but also to express the level of difficulty to detect files that were encoded by such methods. The developed encoders are not an extensive list of what is possible to explore in CPs' data validation tests, but represent extremes of a trade-off between computational effort to encode and detect. Each of the implemented encoders can present *parameters*, *e.g.*, to vary the amount of data, to inject data in different headers, etc. The implemented encoders are explained in the following subsections, sorted by encoder groups. Section IV-D presents the proof of concept system's architecture.

### A. Steganography-related Encoders

Steganography-related encoders use the steganography technique, where messages (or, in this case, data) are hidden in a way that intends to turn them imperceptible apart from the sender and recipient. Steganography was topic of research [10][6] and depending on how it is applied, it can bring a high complexity to be detected. In this paper, the term steganography is applied when data is injected in file's content data (*e.g.*, slightly modifying RGB bits in images, and WAV bits in audio files) that is still not perceivable by humans [11].

**JPG & PNG Steganography Encoder (JPG-PNG-Stega).** The purpose of this encoder is to inject data (hiding it) in JPG and PNG files. It takes a File Format Sample (JPG or PNG) and, for each pixel, it injects 3 bits of data in the LSBs (Least Significant Bit). There are more sophisticated image steganography methods already widely discussed [11]. However, the use of this encoder shows that even basic meth-ods of JPG and PNG steganography are possible in SaaS applications, and the exploitation of such method may bring impacts (Section VI and VII).

**WAV Steganography Encoder (WAV-Stega).** Injecting data in WAV chunk samples [17] is very similar compared to how it is injected in image pixels (as in JPG-PNG-Stega). Based on the WAV file format [14], this encoder checks how many *data samples* (audio samples) there are inside the *data subchunk*. For each sample (that can be composed of *X* Byte, depending on what is declared in *BlockAlign* field), the WAV-Stega encoder injects data in the 4 LSBs of the sample.

**Text Steganography Encoder (TXT-Stega).** The purpose of this encoder is to hide data in a text format, using Steganography. An English dictionary was used to support the TXT-Stega, not requiring a File Format Sample. Based on a sequence of bytes (data that should be hidden), the encoder generates a text output with English words. For testing purposes, the encoder has a pre-defined mapping table with word sets that corresponds to hexadecimal bytes. Since it is a bidirectional function, it is also possible to reverse and get the original input (decoding). Also, this encoder can be adjusted to generate an output of a maximum given number of bytes (*i.e.*, a maximum number of characters).

### B. FileFormatHeader-related Encoders

The FileFormatHeader-related encoders use specific methods depending on the file format. The file format header (or any kind of organizational data) is explored to generate inconsistencies or to inject a high quantity of data. This encoder group is classified with a medium detection complexity, since the fields can be verified iterating throughout the file (considering a standardized file format).

**ID3v2 Tag Encoder (ID3-Tag).** This encoder injects data using optional headers of an audio file format. The ID3-Tag uses the ID3 version 2 metadata container [12] to inject data in some fields related to the audio file, *e.g.*, title, artist, album, track number, among others. The standard specifies a metadata tag size up to 256 MByte. Thus, this encoder uses a File Format Sample up to 10 seconds of audio file and builds a *X* MByte ID3v2 tag within the file. All ID3v2 fields are used by the encoder. As a parameter, it is possible to specify the amount of data to be injected in the ID3v2 fields.

Note that this encoder has the goal to exploit the lack of balance validation between the audio duration, audio quality, and file size. *E.g.*, a MP3 or WAV file with 10 seconds and 256+ MByte cannot be considered coherent, even if generated with the best audio quality.

**JPG Marker Encoder (JPG-Marker).** This encoder takes advantage of additional JPG standardized markers to inject data inside an image file. In the JPG standard [16] is described that the marker "$APP_n$" is application-specific, allowing vendors to store metadata. The marker is represented by the hexadecimal "FF E$n$" (2 Byte), where "$n$" can vary from "2" to "F" (hexadecimal). Therefore, it is possible to have 14 application-specific markers in one JPG file. The "$APP_0$" and "$APP_1$" are reserved to JFIF and EXIF standards, respectively. Each application-specific marker has 2 Byte to store the data length. For this reason, it is possible to have 65,535 Byte of data given the data length representation field. Moreover, there is a marker for textual comments entitled "COMM". This marker is represented by the hexadecimal "FF FE" and can carry the same amount of data as an "APP" (the marker data length representation is also 2 Byte).

The JPG-Marker injects data in these 14 application-specific markers, plus injecting the maximum amount of data in the EXIF marker ("$APP_1$"). It also creates one to many "COMM" markers with data associated to it. The JFIF is left untouched, since the majority of JPG validators fully read it. A JPG file constructed by this encoder contains, at least, 1,048,560 Byte, having 983,025 Byte of the "$APP_{1..15}$" and at least 65,535 Byte of one "COMM" marker. As a parameter, it is possible to add more "COMM" markers for testing purposes.

**PNG Ancillary Chunks Encoder (PNG-Chunks).** The PNG-Chunks encoder injects data in non-mandatory PNG header fields. A chunk is comparable to the JPG marker: it is a data structure expressed after the PNG header [20] that conveys certain information about the image. The chunk consists of four parts: length (4 Byte), chunk type/name (4 Byte), chunk data (length in bytes), and CRC (Cyclic Redundancy Code, 4 Byte).

In order to store data, this encoder uses the chunks entitled "iTXt" and "tEXt". According to the standard both chunks store text, with one name/value pair for each — being possible to have multiple "tEXt" in the same PNG file. In fact, the data representation does not matter since the injected data in these chunks is not displayed.

Due to the 4 Byte length representation, the implemented encoder is able to inject 4 GByte of data in each specified chunk. Therefore, as encoder parameters, it is possible to specify how much data will be injected, and in which name/value pairs (it can present multiple pairs, if needed).

**Email Encoder (Email-Enc).** Email-Enc was implemented to explore the injection of data within the email body and attachments. The implementation is trivial, since it is an enhanced IMAP (Internet Message Access Protocol) email client. The enhanced part that differs from a simple email client is twofold. First, it takes data and transforms in a textual representation to include in the email body (using UTF-8). Second, it also takes data and put as an email attachment. The Email-Enc has two parameters: the amount of data injected in an email body and in the attachment part.

### C. Appending-related Encoder

The Appending-related encoder group appends data in the end of the file in order to check if CPs' data validation checks end-of-file markers. This group is classified with a low detection complexity since the file size can be calculated and end of file markers are defined in standards.

**Appender Encoder (Append-Enc).** Append-Enc takes a File Format Sample as input and appends data after its content data. Since some file formats do not have an explicit end-of-file marker (*e.g.*, MP3), it was created a specific marker to separate the File Format Sample content data to the appended data. The marker is the sequence of bytes "FF FF 01 FF 02 FF 03 FF 04" (hexadecimal). The amount of data to be appended is configurable through a parameter. By default, the Append-Enc appends the same amount of data as contained in the File Format Sample. *E.g.*, if a JPG file has 300 kByte, the encoder appends more 300 kByte of data, resulting in a file of 600 kByte. Based on the appender marker, the decoder is able to retrieve the appended data.

TABLE I
SELECTION OF CLOUD PROVIDERS AND
CLOUD SERVICES, INCLUDING THEIR RESTRICTIONS

| CP | Service | Restrictions |
|---|---|---|
| Google | Picasa | File Formats: JPG, TIFF, BMP, GIF, PSD, PNG, TGA. Images up to 800x800 pixels are not taken into consideration to the free storage (1 GByte). Images beyond 800x800 and up to 2048x2048 are stored taking space within the 1 GByte. File size: up to 20 MByte to upload. |
| Google | GMail | Any text format in the email body. Any File Format as Attachment. The email should be up to 25 MByte. |
| Google | Google+ Status Update | Up to 100,000 characters. |
| Google | Google Docs | Documents: 1,024,000 characters. Uploaded document files that are converted to Google format should be up to 2 MByte. |
| Face-book (FB) | Status Update | 63,206 characters |
| Face-book (FB) | Photos | File formats: GIF, JPG, PNG, PSD, TIFF, JP2, IFF, WBMP, XBM. Up to 720x720 pixels (display). Up to 2048x2048 pixels to upload (but being resized afterwards). Album with a limit of 200 photos. |
| Sound-Cloud | Sound-Cloud Audio | File Formats: WAV, OGG, MP2, MP3, or WMA. Up to *X* minutes of audio. In the free account, up to 120 minutes. |
| TwitPic | Image Service | File Formats: GIF, JPG, and PNG. Images up to 10 MByte to upload. |
| Imgur | Image Service | File Formats: JPG, GIF, PNG, etc. Most files are converted to PNG after the upload. Up to 10 MByte to upload. However, if the image is over 1 Mbyte then it will automatically be compressed or resized to 1 MByte. |
| Image-Shack | Image Service | File Formats: JPG, PNG, ICO, BMP, and TIFF. File size up to 5 MByte to upload (otherwise resized). |

*D. Proof of Concept System*

Fig. 1 depicts the proof of concept system's architecture. The main component is the Test Case Implementation, where the test automation is done. It has the responsibility to prepare a file, push/retrieve to/from the SaaS application, and check if what was pushed is identical to what was retrieved. The File Type Provider has the responsibility to provide file samples to the Test Case Implementation. *E.g.*, if observed that a certain SaaS application only accepts JPG images (Section III-B, Step 1), then the Test Case Implementation requests the File Type Provider a JPG file sample. The File Type Provider selects a pre-defined file out of the specific requested type. The Test Case Implementation interacts with Encoders based on the File Type Sample. The Encoders Interface has two methods: *encode* and *decode*. The encode takes as input the File Type Sample since it needs a starting point to encode data considering a file format. The File Type Sample is not a strictly mandatory input parameter, as it depends on the test case. The output is a file that in terms of structure is very similar to the File Type Sample. However, depending on the Encoder Implementation, the file's organizational and content data might be modified to explore a certain fragility in the data validation process.

The Test Case Implementation also interact with the Cloud Service Interface. This software component is service-specific, and is implemented considering chosen CPs to perform the test cases. For each CP, a Java library that implements push and retrieve of data was used. Thus, the Test Case Implementation delegate the responsibility to push or retrieve the files to specific libraries that directly interact to CSs.

V. TEST CASES

*A. Chosen Cloud Providers*

Table I lists chosen CPs, the selected CSs, and restrictions that matter to encoders and for tests. Note that even if restrictions can change anytime, they were considered to elaborate and perform the test cases. Table I was compiled according to Step 1 of the presented test methodology in Section III-B, and selection criteria in Section III-C.

*B. Test Cases*

All test cases follow the test method presented in Section III-B. Step 2 of the test method describes the construction of data using different encoders. Therefore, test cases in this section intend to specifically describe how the encoder is applied to specific providers with different parameters (*e.g.*, varying the used File Format Samples or the quantity of data injected). Based on the test method, implemented encoders, and chosen providers, the following test cases are described:

1) Encoder: JPG-PNG-Stega. Target: Google Picasa, TwitPic, Facebook (FB) Photos, Imgur, ImageShack. Unique test.
2) Encoder: WAV-Stega. Target: SoundCloud. Unique test.

3) Encoder: TXT-Stega. Target: Facebook Status Update, Google+ Status Update. Unique test.
4) Encoder: ID3-Tag. Target: SoundCloud. Test Cases:
   4.a) ID3v2 tags with 50 MByte of data.
   4.b) ID3v2 tags with 256 MByte of data.
   4.c) ID3v2 tags with 300 MByte of data.
5) Encoder: JPG-Marker. Target: Google Picasa, TwitPic, Facebook Photos, Imgur, ImageShack. All test cases consider the aforementioned 14 "$APP_n$" marker. For each target, the following test cases are performed:
   5.a) JPG with 1 "COMM" marker.
   5.b) JPG with 5 "COMM" markers.
   5.c) JPG with 10 "COMM" markers.
   5.d) JPG with 50 "COMM" markers.
   5.e) JPG with 100 "COMM" markers.
6) Encoder: PNG-Chunks. Target: Google Picasa, TwitPic, Facebook Photos, Imgur, ImageShack. For each target, the following test cases are performed:
   6.a) 1 name/value pair as "tEXt" with 250 kByte, and "iTXt" with 250 kByte of data.
   6.b) 5 name/value pairs as "tEXt" with 250 kByte each, and "iTXt" with 250 kByte of data.
   6.c) 10 name/value pairs as "tEXt" with 250 kByte each, and "iTXt" with 250 kByte of data.
   6.d) 20 name/value pair as "tEXt" with 1 MByte of data each, and "iTXt" with 1 MByte of data.
   6.e) no "tEXt", and "iTXt" with 1 MByte of data.
7) Encoder: Email-Enc. Target: Google GMail. Test cases:
   7.a) Email body: 25 MByte; no attachment.
   7.b) Email body: 25 MByte; Attachment: 1 MByte.
   7.c) Email body: 10 MByte; Attachment: 15 MByte.
   7.d) No email body; Attachment: 5 MByte.
   7.e) No email body; Attachment: 26 MByte.
8) Encoder: Append-Enc. Target: Google Picasa, Google GMail, Google Docs, SoundCloud, TwitPic, Facebook Photos, Imgur, ImageShack. For each target, use the Appender Encoder to:
   8.a) Append the double of the File Format Sample size.
   8.b) Append 1 MByte of data.
   8.c) Append 10 MByte of data.
   8.d) Append 100 MByte of data.

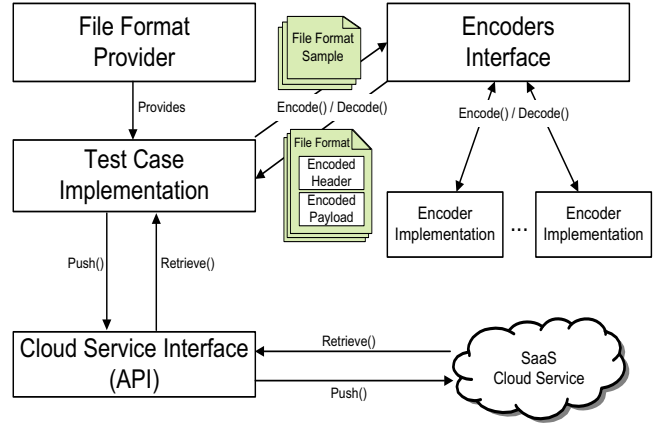

Fig. 1. Proof of Concept System's Architecture.

## VI. RESULTS

The test cases were performed and generated the results shown in Table II, III, and IV. Within table cells, "N/A" stands for "Non-Applicable", meaning that a test case did not target that specific CS; symbol "✓" shows that the test case was *successful*, meaning positive results in test methodology Step 3, 4, 5, **and** 6 (Section III-B); and symbol "✗" means that test methodology Steps 3, 4, 5, **or** 6 were *unsuccessful*.

Test Cases 1, 2, and 3 were successful for the proposed targets. For Test Case 1 should be noted that image file restrictions were respected, since Steganography-related encoders would be unsuccessful due to CPs' image resizing after upload. Test Case 4.c was unsuccessful due to the fact that the pushed MP3 file does not follow ID3v2 standards. The CP did not allow the audio file to be pushed (unsuccessful in Step 4). Test Case 5 demonstrated to have different results depending on the target. While Imgur accepted all files and also made them available to be retrieved in its original form, other CSs denied the upload (unsuccessful in Step 4). The unsuccessful test cases targeting Google Picasa, TwitPic, and Facebook Photos resulted in error messages such as "there was a problem with the image file" or "file must be an image or video". The failed test case targeting ImageShack resulted in a "file too big to be uploaded" error, since the total file size surpassed CS restrictions.

Test Case 6 presented different results due to the following reasons. When Picasa was targeted, it seems that the data validation takes into consideration the existence of ancillary chunks, mainly "iTXt" and "tEXt" (unsuccessful in Step 4). When tried to perform an additional test case, where a PNG file was uploaded with another type of ancillary chunk (containing 1 MByte of data), the test was successful. Facebook accepted to push the encoded files, but when they were retrieved they did not contain the original data (unsuccessful in Step 6). For the other targets, Test Case 6 was unsuccessful when the CS had upload file size restrictions (unsuccessful in Step 4). *E.g.*, ImageShack just accepts 5 MByte, TwitPic 10 MByte, and Imgur 1 MByte. Moreover, Imgur resized the original files after the upload and therefore Test Cases 6.b, 6.c, and 6.d were unsuccessful (Step 6): the retrieved file was not the same as the original.

Within Test Case 7, GMail demonstrated to be coherent to the employed restrictions, not presenting data validation issues. However, combining the Email-Enc with Steganography-, FileHeaderFormat-, and Appender-related encoders, it was possible to retrieve hidden information persisted in email body and attachment files.

Test Case 8 demonstrated to be successful even though it is the most trivial encoder to detect and prevent. For Picasa, GMail, Docs, TwitPic, Imgur, and ImageShack, Test Cases 8.d and 8.c was unsuccessful due to the upload file size restriction and not due to the used encoder. Thus, the encoded file was not even accepted for upload (unsuccessful in Step 4). Facebook Photos accepted to upload the files. But within Step 5 and 6 the data validation transformed the image files and cut the appended data.

## VII. Discussion

### A. Security Impacts

The performed tests identified a considerable security impact: unawareness of persisted content. This is related to security since CPs may distribute data (*e.g.*, by sharing a picture, or turning public an audio file) that they do not filter and are not aware of. CPs may face legal issues due to the distribution of, *e.g.*, illegal content under the jurisdiction where the data is persisted. Nowadays CPs have multiple datacenter facilities around the globe, employing a Content Delivery Network (CDN) to distribute/replicate data where is appropriated (based on, *e.g.*, content rules). If these content rules do not take the injected data into consideration, CDNs are prone to take unappropriated distribution/replication actions. A typical example to illustrate this impact can be observed if a user persists copyrighted software into a CP and publicly share it — assuming people know how to decode and get the injected data out of the encoded file. Malicious users can benefit from CP unawareness to build, *e.g.*, a peer-to-peer software that is able to encode, persist in different CPs, share, and decode any kind of data, independent of content type.

As long as data validation processes remain ill-defined, and malicious users exploit it in a large scale, other security impacts as application failures and Denial-of-Service may be observed. *E.g.*, after the data validation, the SaaS application may process the file to display in the Website, or add functions as image editing. If the whole process is not consistent, the CS can suffer faults due to unexpected organizational and content data that bypassed the data validation process.

### B. Accounting and Charging Impacts

In two test cases, the chosen CPs suffered impacts in their accounting and charging systems. Google Picasa provides a free account with 1 GByte of storage. Table I shows that "images up to 800x800 pixels are *not* taken into consideration to the free storage (1 GByte), and images from 800x800 up to 2048x2048 pixels are stored consuming space within the 1 GByte space". The Test Cases 1, 5, and 8 used images with a resolution below 800x800, injecting data, but not enhancing the File Format Sample resolution. After uploading the encoded image files, the account space consumption remained with 1 GByte, even if the encoded images were very large. This means that, by exploiting the data validation process, users are able to store any data without being accounted and charged for — since the free 1 GByte of storage remains unused.

SoundCloud provides a free account with 120 minutes with audio storage. Test Case 4 is able to store MP3 audio files with 10 seconds and 256 MByte. An MP3 file, with sample rate of 44,100 Hz, bit rate of 128 kbit/s, no ID3v2 tags, and 10 seconds, consumes on average 150 kByte. A free account is intended to store 720 MP3 files with 10 seconds each, consuming ~105 MByte; however, by employing the ID3-Tag encoder, it was possible to consume up to 180 GByte with the same amount of files.

TABLE II
RESULTS OF TEST CASE 8.

| Test Cases | | Cloud Service | | | | | | | |
|---|---|---|---|---|---|---|---|---|---|
| | | Google Picasa | Google Docs | Google GMail | Sound Cloud | TwitPic | FB Photos | Imgur | Image Shack |
| 8 | a | ✓ | ✓ | ✓ | ✓ | ✓ | ✗ | ✓ | ✓ |
| | b | ✓ | ✓ | ✓ | ✓ | ✓ | ✗ | ✓ | ✓ |
| | c | ✓ | ✗ | ✓ | ✓ | ✓ | ✗ | ✗ | ✓ |
| | d | ✗ | ✗ | ✗ | ✓ | ✗ | ✗ | ✗ | ✗ |

In both cases a considerable additional amount of data may be stored in CPs' servers without additional charges — since data volume is not accounted and not considered by the data validation process.

## VIII. Conclusions

The carried exploratory tests introduce a novel problem in the area of CC, which could be observed in several on-production SaaS applications. The results showed that, depending on the CP, CS, and used encoder, it is possible to store arbitrary data into the investigated CPs, bypassing the data validation process, thus causing impacts in security, accounting, and charging.

Surprisingly, even non-sophisticated techniques as the ones implemented in FileFormatHeader- and Appender-related encoders demonstrated to be successful. CPs should concentrate their efforts enhancing data validation algorithms to detect the use of encoders that are easier to circumvent but can lead to a larger amount of data being stored — compared to steganography, that requires a high complexity to detect. Even though the use of steganography can always bypass CPs' data validation, the amount of injected data was not as high as when using other encoders.

Within accounting and charging impacts, this paper has identified a novel dependency relation between application-specific accounting attributes (*e.g.*, image resolution, audio length) and what the data validation process takes into consideration. CPs should verify whether values declared in headers reasonably correspond to the amount of resources consumed. *E.g.*, can a 10 second-long MP3 file, with the best possible audio quality, have 256+ MByte if a considerable amount of data is not present in optional headers? Even though data validation rules should follow standards (thus, being possible, *e.g.*, to use 256 MByte just for ID3v2 tags), absurd imbalances between what is declared and what is consumed should be taken into consideration.

As future work, it is important to understand the complexity of preventing usage of the encoders classified in this paper. Finally, the possibility and scalability of building a P2P system to store and share arbitrary data from SaaS applications shall be investigated.

TABLE III
RESULTS OF TEST CASES 1, 5, AND 6.

| Test Cases | | Cloud Service | | | | |
|---|---|---|---|---|---|---|
| | | Google Picasa | TwitPic | FB Photos | Imgur | Image Shack |
| 1 | | ✓ | ✓ | ✓ | ✓ | ✓ |
| 5 | a | ✓ | ✓ | ✗ | ✓ | ✓ |
| | b | ✓ | ✓ | ✗ | ✓ | ✓ |
| | c | ✓ | ✓ | ✗ | ✓ | ✓ |
| | d | ✗ | ✗ | ✗ | ✓ | ✓ |
| | e | ✗ | ✗ | ✗ | ✓ | ✗ |
| 6 | a | ✗ | ✓ | ✗ | ✓ | ✓ |
| | b | ✗ | ✓ | ✗ | ✗ | ✓ |
| | c | ✗ | ✓ | ✗ | ✗ | ✓ |
| | d | ✗ | ✗ | ✗ | ✗ | ✗ |
| | e | ✗ | ✓ | ✗ | ✓ | ✓ |

TABLE IV
RESULTS OF TEST CASES 2, 3, 4 AND 7.

| Test Cases | | Cloud Service | | | |
|---|---|---|---|---|---|
| | | Sound Cloud | Google GMail | Google+ Status Update | FB Status Update |
| 2 | | ✓ | N/A | N/A | N/A |
| 3 | | N/A | N/A | ✓ | ✓ |
| 4 | a | ✓ | N/A | N/A | N/A |
| | b | ✓ | N/A | N/A | N/A |
| | c | ✗ | N/A | N/A | N/A |
| 7 | a | N/A | ✓ | N/A | N/A |
| | b | N/A | ✗ | N/A | N/A |
| | c | N/A | ✓ | N/A | N/A |
| | d | N/A | ✓ | N/A | N/A |
| | e | N/A | ✗ | N/A | N/A |